\documentclass[11pt,a4paper]{article}

\usepackage[utf8]{inputenc}
\usepackage{bold-extra}
\usepackage{chngcntr}
\usepackage{tikz}
\usepackage{setspace}
\usepackage{url}
\usepackage{amsfonts}
\usepackage{amssymb}
\usepackage{epsfig}
\usepackage{amsmath}
\usepackage{commath}
\usepackage{natbib}
\usepackage[linesnumbered,ruled]{algorithm2e}
\usepackage{graphicx}
\usepackage{bbm,float}
\usepackage{dsfont}
\usepackage{color}
\usepackage{adjustbox}
\usepackage{amsmath,amssymb,amstext,amsfonts,amsthm,latexsym}
\usepackage{diagbox}
\usepackage{bm}
\usepackage{caption}
\usepackage{subcaption}
\usepackage{multirow}
\usepackage{color}
\usepackage{listings}
\usepackage{titlesec}
\usepackage{enumerate}
\usepackage{environ}
\usepackage{pdflscape}
\usepackage{booktabs}
\usepackage{appendix}
\usepackage{bbm}
\usepackage{pgfplots}
\usepackage{authblk}
\usepackage{makecell}
\usepackage{tabularx}
\usepackage{blkarray}
\newcolumntype{Y}{>{\centering\arraybackslash}X}
\numberwithin{equation}{section}
\theoremstyle{plain}

\newtheorem{proposition}{Proposition}[section]

\theoremstyle{definition}

\newtheorem{definition}{Definition}[section]
\newtheorem{example}{Example}[section]

\theoremstyle{remark}

\pgfplotsset{compat=1.17}

\definecolor{darkblue}{rgb}{0.0, 0.0, 0.55}
\newcommand{\EF}[1]{\textcolor{darkblue}{#1}}

\begin{document}
\title{\textbf{Mean-tail Gini framework for optimal portfolio selection}
}
\date{\today}
\author[1,2]{Jinghui Chen\thanks{Corresponding author: \texttt{jh8chen@yorku.ca}}}
\author[1,2]{Edward Furman}
\author[1]{Stephano Ricci}
\author[1]{Judeto Shanthirajah}

\affil[1]{Department of Mathematics and Statistics at York University}
\affil[2]{RISC Foundation}
\maketitle
\begin{abstract}
	
	The limitations of the traditional mean-variance (MV) efficient frontier, as introduced by \cite{markowitz1952portfolio}, have been extensively documented in the literature. Specifically, the assumptions of normally distributed returns or quadratic investor preferences are often unrealistic in practice. Moreover, variance is not always an appropriate risk measure, particularly for heavy-tailed and highly volatile distributions, such as those observed in insurance claims and cryptocurrency markets, which may exhibit infinite variance. To address these issues, \cite{shalit2005mean} proposed a mean-Gini (MG) framework for portfolio selection, which requires only finite first moments and accommodates non-normal return distributions.
	
	However, downside risk measures - such as tail variance - are generally considered more appropriate for capturing risk managers' risk preference than symmetric measures like variance or Gini. In response, we introduce a novel portfolio optimization framework based on a downside risk metric: the tail Gini. In the first part of the paper, we develop the mean-tail Gini (MTG) efficient frontier. Under the assumption of left-tail exchangeability, we derive closed-form solutions for the optimal portfolio weights corresponding to given expected returns. In the second part, we conduct an empirical study of the mean-tail variance (MTV) and MTG frontiers using data from equity and cryptocurrency markets. By fitting the empirical data to a generalized Pareto distribution, the estimated tail indices provide evidence of infinite-variance distributions in the cryptocurrency market. Additionally, the MTG approach demonstrates superior performance over MTV strategy by mitigating the amplification distortions induced by $\mathrm{L}^2$-norm risk measures. The MTG framework helps avoid overly aggressive investment strategies, thereby reducing exposure to unforeseen losses.
	
\end{abstract}


\newpage
\section{Introduction}
Modern portfolio theory has long been anchored in the MV model introduced by \cite{markowitz1952portfolio}. Despite its widespread preference in both academia and the financial industry, the MV framework relies on restrictive assumptions - most notably, the normality of asset return distributions and the finiteness of their second moments - that often fail to hold in practice. Empirical evidence provided by \cite{chunhachinda1997portfolio}, for instance, reveals significant deviations from normality in returns from major global equity markets. These shortcomings become particularly pronounced in markets such as insurance and cryptocurrency, where distributions frequently exhibit heavy tails and significant skewness. \cite{chen2024infinite} discussed that power-law models with infinite second moments are ubiquitous, and emphasized the usefulness of infinite-moment models in risk management, such as modeling catastrophic and operational risks. For example, \cite{finma2021} reported empirically that the default tail parameter $\alpha$ of power-law models for most major damage insurance losses is in the range of $[1,2]$, which means the variance is infinite. Moreover, the possibility of using models with infinite moments cannot be excluded due to the finiteness of sample moments \citep{mandelbrot2013fractals}. \cite{denneberg1990premium} and \cite{konno1991mean} explained that variance distorts (amplify) deviations due to the $\mathrm{L}^2$ norm and hence is not great for heavy-tailed distributions of returns.   

To address these deficiencies, alternative portfolio selection models have been proposed. Renowned examples include the three-moment model \citep{simaan1993portfolio}, four-moment framework \citep{jondeau2006optimal}, the value-at-risk (VaR) based approach \citep{campbell2001optimal,basak2001value}, as well as the conditional value-at-risk (CVaR) risk measure \citep{cai2024worst,cai2025conditional}. Among these, the mean-Gini (MG) framework developed by \cite{shalit1984mean, shalit2005mean} offers a robust alternative by replacing the MV model's dependence on quadratic utility and normality with the Gini mean difference (GMD) as a measure of risk. For a random variable (RV) $L$ with cumulative distribution function (CDF) $F$, the GMD, denoted by $G_L$, is defined as
\begin{equation*}
	G_L=\mathrm{E}[\lvert L_1-L_2\rvert]=4\mathrm{Cov}[L, F(L)],
\end{equation*} where RVs $L_1$ and $L_2$ are two independent copies with CDF $F$, $\mathrm{E}$ and $\mathrm{Cov}$ denote the expected value of a RV and the covariance of two RVs, respectively. In contrast, the variance of $L$ is given by
\begin{equation*}
	\sigma_L^2=\frac{1}{2}\mathrm{E}[(L_1-L_2)^2]=\mathrm{Cov}[L, L].
\end{equation*}Interestingly, variance and GMD exhibit several conceptual and mathematical similarities, as both serve as measures of statistical dispersion. For a comprehensive discussion of their comparative properties, we refer the reader to Section~2 of \cite{yitzhaki2003gini}. Although both variance and GMD quantify dispersion, they are symmetric measures and, therefore, fail to isolate downside risk - an aspect of particular importance to investors and risk managers, for whom under-performance is of greater concern than over-performance.

In response, \cite{estrada2002systematic,estrada2007mean,estrada2008mean} introduced the mean-semi-variance (MSV) framework, which minimizes the semi-variance - a measure that focuses solely on returns below a specified threshold $B$. The semi-variance of a portfolio return $L$, denoted by $\Sigma_B^2$, is defined as
\begin{equation*}
	\Sigma_B^2=\mathrm{E}[(\min(L-B, 0))^2]=\mathrm{E}[(L-B)^2\mathds{1}_{L<B}],
\end{equation*}where $B$ is a threshold parameter chosen by the risk manager. Additionally, \cite{furman2006tail} and \cite{furman2017gini} introduced the tail variance risk measure in actuarial science, denoted TV$_{p}(L)$ for prudence levels $p\in(0,1)$, as
\begin{equation*}
\mathrm{TV}_{p}(L)=\mathrm{E}[(L-\mathrm{TCE}_{p}(L))^2\vert L<\mathrm{VaR}_{p}(L)],
\end{equation*}
where $\mathrm{TCE}_{p}(L)=\mathrm{E}[L\vert L<\mathrm{VaR}_{p}(L)]$ denotes the tail conditional expectation. \cite{landsman2010tail} then generalized the MV framework to the mean-tail variance (MTV) approach by minimizing the risk measure $\mathrm{TV}_{p}(L)$. \cite{owadally2013characterization} further discovered the characterization of optimal portfolios obtained via the MTV. We refer readers to \cite{eini2021tail} and \cite{huang2024tail} for more research in this topic.

While the MSV and MTV models explicitly accounts for downside risk, them, like the MV framework, presupposes the existence of finite second moments. However, such assumption is often violated in the case of insurance claims and cryptocurrency returns, where infinite variance may arise. For instance, \cite{dhaene2002concept} argue that natural catastrophic risks may exhibit infinite second moments, while \cite{grobys2023fractal} provides empirical evidence that the infinite variance hypothesis cannot be rejected for Bitcoin and S\&P 500 returns - a finding aligned with the classic study on cotton price fluctuations from \cite{mandelbrot1963variation}.  

Building on these developments, we propose a novel framework: mean-tail Gini (MTG) portfolio optimization. The MTG approach employs tail Gini - a measure that emphasizes the left tail of the return distribution, i.e., downside risk - as the risk criterion. Following \cite{furman2017gini}, the tail Gini of a return variable $L\sim F$ at a prudence level $p\in(0,1)$, denoted by $\mathrm{TGini}_p(L)$, is defined as
\begin{equation*}
	\mathrm{TGini}_p(L) = \frac{4}{p} \mathrm{Cov}[L, F(L) | L<\mathrm{VaR}_p(L)],
\end{equation*} where $\mathrm{VaR}_p(L)=\inf\{l\in\mathbb{R}: F(l)\geq p\}$ is the VaR of $L$ at the prudence level $p$. The MTG framework minimizes the tail Gini to construct an efficient frontier analogous to those generated by the MV and MSV models, but with a sharper focus on downside risk and without the restrictive assumption of finite variance. It is evident  that the tail Gini is positive as it is the conditional expectation of absolute value of difference between two independent variables.
 	
To derive closed-form expressions for optimal portfolios under MTG, we propose the concept of tail exchangeability in the first part of the paper. A bivariate return vector $(L_1,L_2)$ is said to be tail exchangeable if the tail Gini correlation coefficients between $L_1$ and $L_2$ are symmetric. Assuming that all pairs of constituent asset returns, as well as each return and the portfolio return, are tail exchangeable, the MTG efficient frontier can be derived analytically by replacing the Pearson correlation matrix used in the MV approach with its tail Gini counterpart. Moreover, this concept can also be employed to detect deviations from normality in return distributions.

In the second part of the paper, we empirically examine the MTG efficient frontier using daily returns from six asset classes, including stocks, bonds, and cryptocurrencies, in the U.S. market. By fitting the empirical data to a generalized Pareto distribution, the tail indices demonstrate that infinite variance distributions do exist in the cryptocurrency market. Our analysis shows that when short selling is permitted, the MTG framework outperforms the MV approach by delivering a higher expected return for a given level of risk. Conversely, in the absence of short selling, MV portfolios tend to incorporate more assets than prudence-based measures. Additionally, our empirical findings demonstrate that short selling is a valuable tool for enhancing market efficiency, which is consistent with the findings in the literature; see, e.g., \cite{saffi2011price}. Moreover, the MTG approach demonstrates better performance over the MTV strategy by mitigating the amplification distortions induced by $\mathrm{L}^2$-norm risk measures.

The remainder of the paper is organized as follows. Section~\ref{MTG framework} introduces the MTG framework and derives the analytical efficient frontier under the tail exchangeability assumption. Section~\ref{data desc and params calib} describes the data and calibration of parameters for the six asset classes. Section~\ref{empirical MTG EF} presents the empirical analysis and comparisons across the MTG, MV, and mean-tail variance (MTV) models. Finally, Section~\ref{conclusion} concludes the paper.

\section{Mean-tail Gini framework}\label{MTG framework}
\subsection{Definition and problem setting}
Let \((\Omega, \mathcal{A}, \mathbb{P})\) denote a standard atomless probability space, and let $L^0 := L^0(\Omega, \mathcal{A}, \mathbb{P})$ represent the set of all real-valued RVs $L,\ L_i,\ i\in\mathcal{N}=\{1,2,\dots,d\},\ d \in \mathbb{N}$ on this probability space. Throughout this paper, the notation $L\sim F_L$ indicates that the RV $L$ follows the CDF \(F_L(l) = \mathbb{P}(L \leq l), \, l \in \mathbb{R}\). The function \(F_L^{-1}(p)\) for $p \in (0, 1)$ denotes the generalized quantile function of \(F_L\) defined as  
\[
F_L^{-1}(p) = \inf \{l \in \mathbb{R} : F_L(l) \geq p \}.
\] Unless otherwise specified, we assume that all RVs involved have finite expectations, and $U$ is a standard uniformly distributed RV. 

Let us consider a portfolio where the return is modeled by the RV  $$L=\bm{\alpha}'\bm{L}=\sum_{i=1}^{d}\alpha_iL_i\sim F,$$ with $\bm{L}=(L_1,L_2,\dots,L_d)'$ representing a $d$-dimensional random vector of asset returns, and $\bm{\alpha}=(\alpha_1,\alpha_2,\dots,\alpha_d)'$ denoting the vector of portfolio weights. In modern portfolio theory, risk managers seek to minimize a specific risk measure of the portfolio return $L$, denoted by $\rho(L)$, subject to a predetermined expected value. Examples of $\rho(L)$ are VaR \citep{campbell2001optimal} and CVaR \citep{cai2024worst}. The efficient frontier and optimal portfolio weights $\bm{\alpha}^*$ are then characterized by the solution to the following optimization problem:
\begin{equation}\label{optim problem}
	\begin{aligned}
		\min\ &\rho(L) \\
		\text{subject to } &\mu = \bm{\alpha}'\bm{\mu}\\
		&1=\bm{\alpha}'\bm{1},
	\end{aligned}
\end{equation} where $\bm{\mu}=(\mu_1,\mu_2,\dots, \mu_d)'=(\mathrm{E}[L_1],\mathrm{E}[L_2],\dots, \mathrm{E}[L_d])'$ denotes the vector of expected returns, and $\bm{1}=(1,1,\dots,1)'$ is a $d$-dimensional vector of ones.

The most well-known risk measure in the literature and industry is the standard deviation of the MV framework. In such a case, the risk measure in optimization problem \eqref{optim problem} is the variance defined as the following:
\begin{equation*}
		\rho(L) :=\sigma_L^2=\bm{\alpha}'\bm{\Sigma}\bm{\alpha}=\bm{\alpha}'\bm{A}\bm{R}\bm{A}\bm{\alpha},
\end{equation*}where $\bm{\Sigma}$ and $\bm{R}$ denote the $d\times d$ covariance and Pearson correlation coefficient matrices of the random vector $\bm{L}$, respectively, and $\bm{A}$ is a diagonal $d\times d$ matrix with the standard deviation $\sigma_i$ of RV $L_i$ as diagonal entries and zeros elsewhere. Specifically, $\bm{R}$ consists of the correlation coefficients of $L_i$ and $L_j$ in its $i$-th row and $j$-th column, denoted by,
\begin{equation*}
	\rho_{ij}=\frac{\mathrm{Cov}(L_i,L_j)}{\sigma_i\sigma_j}.
\end{equation*}

It is evident that the risk measure $\sigma_L^2$ is not universally effective. For example, if the portfolio incorporate one asset such that its return $L_1$ follows the generalized Pareto distribution with a shape parameter in $[0.5, 1)$, $\sigma_L^2$ is infinite with any $\bm{\alpha}$, and $\rho_{i1}=0$. \cite{chen2024infinite} emphasized the importance of infinite second moments in modeling catastrophic and operational risks. Furthermore, as an $\mathrm{L}^2$-norm risk measure, $\sigma_L^2$ disproportionately amplifies deviations from the mean, which is particularly undesirable when dealing with heavy-tailed distributions \citep{konno1991mean}. Moreover, $\rho_{ij}$ is not an appropriate dependence measure since it is affected by marginal distributions. For a detailed discussion on the shortcomings of Pearson correlation, we refer to \citet{embrechts2002correlation}. 

To address certain limitations of the variance, \cite{shalit1984mean,shalit2005mean} proposed the MG framework. In response to Pearson correlation coefficient, C.\ Gini proposed an alternative dependence measure over a century ago, now widely known as the Gini correlation coefficient; see, for example, \cite{yitzhaki2013gini} and \cite{furman2017beyond}. The following definition introduces the concept of Gini correlation coefficients.
\begin{definition}[Gini correlation coefficients]\label{gini corr coef}
	Let $L_1\sim F_1$ and $L_2\sim F_2$	be two RVs. The Gini correlation coefficients are given by
	\begin{equation*}
		\Gamma_{12}=\frac{\mathrm{Cov}[L_1,F_2(L_2)]}{\mathrm{Cov}[L_1,F_1(L_1)]}\text{ and } \Gamma_{21}=\frac{\mathrm{Cov}[L_2,F_1(L_1)]}{\mathrm{Cov}[L_2,F_2(L_2)]}.
	\end{equation*}
\end{definition}We refer readers to \cite{yitzhaki2003gini} for the detailed properties of Gini correlation coefficients. Let $G_i$ be the GMD of $L_i$ and $\Gamma_{iL}$ be the Gini correlation coefficient between $L_i$ and $L$ for $i\in\mathcal{N}$. In MG framework, the risk measure in optimization problem \eqref{optim problem} is the GMD defined as the following:
\begin{equation*}
\rho(L) :=G_L=4\bm{\alpha}'\bm{K}\overset{\Gamma_{ij}=\Gamma_{ji}}{=}\sqrt{\bm{\alpha}'\bm{B}\bm{\Gamma}\bm{B} \bm{\alpha}},
\end{equation*}where $\bm{K}=(G_1\Gamma_{1L}, G_2\Gamma_{2L},\dots, G_d\Gamma_{dL})'$ is a vector includes covariances between $L_i$ and $F(L)$, $\bm{A}$ is a diagonal $d\times d$ matrix with $G_i$ as diagonal entries and zeros elsewhere, and $\bm{\Gamma}$ is a $d\times d$ matrix consists of $\Gamma_{ij}$ in its $i$-th row and $j$-th column. \cite{yitzhaki2003gini} proved the last equality under the assumption of $\Gamma_{ij}=\Gamma_{ji}$.

In this paper, inspired by the movement from variance to semi-variance, we propose a MTG portfolio optimization framework, which minimize the tail Gini, as a tail-focused analogue to the GMD. Similarly, we first introduce a tail dependence measure, named the tail Gini correlation coefficient. We refer readers to \cite{liu2021theory} and \cite{landsman2005tail} for an axiomatic study and an examination of the exponential dispersion models for tail risk measures, respectively. 

\begin{definition}[Tail Gini correlation coefficients]
	Let $L_1\sim F_1$ and $L_2\sim F_2$	be two RVs. The tail Gini correlation coefficients at a prudence level of $p\in(0,1)$ are given by
	\begin{equation*}
		\Gamma_{12,p}=\frac{\mathrm{Cov}[L_1,F_2(L_2)| L_2<l_{p,2}]}{\mathrm{Cov}[L_1,F_1(L_1)| L_1<l_{p,1}]}\text{ and } \Gamma_{21,p}=\frac{\mathrm{Cov}[L_2,F_1(L_1)| L_1<l_{p,1}]}{\mathrm{Cov}[L_2,F_2(L_2)| L_2<l_{p,2}]},
	\end{equation*} where $l_{p,1}=\mathrm{VaR}_p(L_1)$ and $l_{p,2}=\mathrm{VaR}_p(L_2)$.
\end{definition}It is evident that when $p$ goes to one, the tail Gini correlation coefficients become the classical Gini correlation coefficients, i.e., $\lim\limits_{p\rightarrow 1}\Gamma_{12,p}=\Gamma_{12}$ and $\lim\limits_{p\rightarrow 1}\Gamma_{21,p}=\Gamma_{21}$. Additionally, the tail Gini correlations are special cases of the weighted Gini correlations; see, e.g., \cite{furman2017beyond}. We refer the reader to \cite{furman2015paths} for copulas-based tail dependence measurement.

The tail Gini of the portfolio return $L$ at a prudence level $p$ is given by
\begin{equation}\label{tail gini eq}
	\begin{aligned}
		\mathrm{TGini}_p(L) &= \frac{4}{p} \mathrm{Cov}[L, F(L) | L<l_p]\\
		&= \frac{4}{p} \sum_{i=1}^{d}\alpha_i\mathrm{Cov}[L_i, F(L) | L<l_p]\\
		&=\frac{4}{p}\bm{\alpha}'\bm{T},
	\end{aligned}
\end{equation} where $l_p=\mathrm{VaR}_p(L)$, and the vector $$\bm{T}=(\mathrm{TGini}_p(L_1)\Gamma_{1L,p},\mathrm{TGini}_p(L_2)\Gamma_{2L,p},\dots, \mathrm{TGini}_p(L_d)\Gamma_{dL,p})'$$ captures the conditional covariances between $L_i$ and $F(L)$, where $\Gamma_{iL,p}$ are the tail Gini correlation coefficient of $L_i$ and $L$. 

In the MTG model, risk managers seek to minimize the tail Gini of a portfolio return, subject to a predetermined expected value. Thus, the optimization problem \eqref{optim problem} is specified as
\begin{equation}\label{MTG opt prob}
	\begin{aligned}
		\min\ &\frac{4}{p}\bm{\alpha}'\bm{T}, \\
		\text{subject to } &\mu = \bm{\alpha}'\bm{\mu},\\
		&1=\bm{\alpha}'\bm{1}.
	\end{aligned}
\end{equation} 

In general, an analytic solution for the optimal portfolio weights $\bm{\alpha}^*$ corresponding to the optimization problem~\eqref{MTG opt prob} is not available. However, under the assumption of tail exchangeability, introduced in Section~\ref{tail exc}, an explicit solution can be derived (Section~\ref{analytic solution}).

\subsection{The tail exchangeability}\label{tail exc}
\cite{wang2016joint} introduced the concept of joint mixability for the d-tuple of probability distributions ($F_1,F_2,\dots,F_d$) if there exist $L_i\sim F_i$, $i\in\mathcal{N}$, such that $L_1+L_2+\cdots +L_d$ is almost surely constant. Motivated by this idea, we define a class of bivariate distributions with symmetric tail dependence, referred to as tail exchangeable distributions, as follows.

\begin{definition}[Tail exchangeability]
	A pair of CDFs ($F_1,F_2$) on $\mathbb{R}$ is left-tail exchangeable if there exist RVs $L_1\sim F_1$ and $L_2\sim F_2$ and $p^\ast\in(0,\ 1)$, such that $\Gamma_{12,p}=\Gamma_{21,p}$ for all $p\in (0,p^\ast]$.
\end{definition}
The left-tail exchangeable $(F_1,F_2)$ possesses the following properties.
\begin{proposition}[Invariance under affine transformations] Let $L_1\sim F_1$ and $L_2\sim F_2$. 
	\begin{enumerate}[(1)]
		\item If the joint distribution of $(L_1,L_2)$ is left-tail exchangeable, then for any constants $a_i\in\mathbb{R}$ and $b_i>0$, $i=1,2$, the joint distribution of $(a_1 + b_1L_1, a_2 + b_2L_2)$ is also left-tail exchangeable.
		\item Suppose $L_i=F_i^{-1}(U_i)$, $i=1,2,$ where $U_i\sim U[0,1]$ are comonotonic on $(0,p^\ast]$, then the joint distribution of $(L_1, L_2)$ is left-tail exchangeable.
		\item Suppose $L_i=F_i^{-1}(U_i)$, $i=1,2,$ where $U_i\sim U[0,1]$ are antimonotonic on $(0,p^\ast]$, then the joint distribution of $(L_1, L_2)$ is left-tail exchangeable.
	\end{enumerate}
	\label{basic p of te}
\end{proposition}

We next present some examples illustrating left-tail exchangeability in a commonly used multivariate model.
\begin{proposition}\label{normal tail exch}
	Let $\bm{L}=(L_1,L_2,\dots,L_d)'$ be a $d$-dimensional Gaussian random vector. Then, the joint distribution of $(L_i,L_j)$ is left-tail exchangeable for all $i,j\in\mathcal{N}$.
\end{proposition}

Proposition~\ref{normal tail exch} confirms the multivariate Gaussian distribution is left-tail exchangeable. The same reasoning extends to other multivariate distributions belonged to the location-scale family, indicating that left-tail exchangeability holds more generally.

\begin{example} Let $L_1\sim F_1$ and $L_2\sim F_2$ denote two discrete RVs having, respectively, five mass points $\{-2,-1,0,1,2\}$ and $\{-1,1,3,5,7\}$ with equal probability. Suppose $L_1$ and $L_2$ follow the following dependence structure
\begin{equation*}
	\begin{array}{c}
		\begin{array}{cc}
			L_1 & L_2
		\end{array} \\
		\begin{bmatrix}
			-2 & -1 \\
			-1 & 1 \\
			0 & 7 \\
			1 & 3 \\
			2 & 5
		\end{bmatrix}
	\end{array}.
\end{equation*}
When $p^\ast< \frac{3}{5}$, we have $\Gamma_{12,p}=\Gamma_{21,p}=1$ for any $p\in(0,p^\ast]$ (See (2) of Proposition~\ref{basic p of te}). In this case, the joint distribution of $(L_1, L_2)$ is left-tail exchangeable. However, if $p^\ast\geq \frac{3}{5}$, such a statement is not true. For example, when $p^\ast= \frac{4}{5}$,
	\begin{equation*}
		\begin{aligned}
			\Gamma_{12,\frac{4}{5}}&=\frac{\mathrm{Cov}[L_1,F_2(L_2)| L_2<5]}{\mathrm{Cov}[L_1,F_1(L_1)| L_1<1]}=\frac{\mathrm{E}[L_1F_2(L_2)| L_2<5]-\mathrm{E}[L_1| L_2<5]\mathrm{E}[F_2(L_2)| L_2<5]}{\mathrm{E}[L_1F_1(L_1)| L_1<1]-\mathrm{E}[L_1| L_1<1]\mathrm{E}[F_1(L_1)| L_1<1]}\\
			&=\frac{-\frac{1}{15}-(-\frac{2}{3}\times \frac{2}{5})}{-\frac{4}{15}-(-1\times \frac{2}{5})}=\frac{3}{2},
		\end{aligned}
	\end{equation*} and 
	\begin{equation*}
		\begin{aligned}
			\Gamma_{21,\frac{4}{5}}&=\frac{\mathrm{Cov}[L_2,F_1(L_1)| L_1<1]}{\mathrm{Cov}[L_2,F_2(L_2)| L_2<5]}=\frac{\mathrm{E}[L_2F_1(L_1)| L_1<1]-\mathrm{E}[L_2| L_1<1]\mathrm{E}[F_1(L_1)| L_1<1]}{\mathrm{E}[L_2F_2(L_2)| L_2<5]-\mathrm{E}[L_2| L_2<5]\mathrm{E}[F_2(L_2)| L_2<5]}\\
			&=\frac{\frac{22}{15}-(\frac{2}{5}\times \frac{7}{3})}{\frac{10}{15}-(1\times \frac{2}{5})}=2.
		\end{aligned}
	\end{equation*} $\Gamma_{12,\frac{4}{5}}\neq \Gamma_{21,\frac{4}{5}}$ supports the previous point.
	\label{example of te}
\end{example}

\subsection{Analytical derivation of the MTG efficient frontier}\label{analytic solution}
In this subsection, we analytically derive the MTG efficient frontier using an approach analogous to that employed for the MV framework. For a detailed derivation of the MV efficient frontier, We refer to \cite{merton1972analytic} and \cite{huang1988foundations}. The proposition below is a generalization of Proposition~2 of \cite{yitzhaki2003gini}. When $p$ goes to one, our Proposition~\ref{tail gini equalvalent to gini} is consistent with his Proposition~2.
\begin{proposition}\label{tail gini equalvalent to gini} Let $\bm{L}=(L_1,L_2,\dots,L_d)'$ follow a multivariate distribution with finite mean vector $\bm{\mu}=(\mu_1,\mu_2,\dots,\mu_d)'$, and define the portfolio return as $L = \sum_{i=1}^{d}\alpha_iL_i\sim F$, with $l_p=\mathrm{VaR}_p(L)$. Then,
	\begin{equation}\label{tail gini}
		\begin{aligned}
			(\mathrm{TGini}_p(L))^2&-\mathrm{TGini}_p(L)\sum_{i=1}^{n}\alpha_iD_{iL, p}\mathrm{TGini}_p(L_i)\\&=\frac{1}{2}\sum_{i=1}^{n}\sum_{j=1}^{n}\alpha_i\alpha_j\mathrm{TGini}_p(L_i)\mathrm{TGini}_p(L_j)(\Gamma_{ij,p}+\Gamma_{ji,p}),
		\end{aligned}
	\end{equation} where $D_{iL, p}=\Gamma_{iL,p}-\Gamma_{Li,p}$ for $i\in\mathcal{N}$.
	When $p$ goes to zero in Equation~\eqref{tail gini}, we obtain
	\begin{equation}\label{gini}
		G^2_L-G_L\sum_{i=1}^{n}\alpha_iD_{iL}G_{L_i}=\frac{1}{2}\sum_{i=1}^{n}\sum_{j=1}^{n}\alpha_i\alpha_jG_{L_i}G_{L_j}(\Gamma_{ij}+\Gamma_{ji}),
	\end{equation}	
	where $G_Y$ denotes the GMD of a RV $Y$.
\end{proposition}

Hereafter, we illustrate a specific solution of the MTG efficient frontier. Under the assumption of left-tail exchangeability, i.e., symmetry in tail Gini correlation coefficients among all pairs $(L_i,L_j)$ and $(L_i,L)$, the efficient frontier can be derived by simplifying Equation~\eqref{tail gini}. For example, when $\bm{L}$ follows a multivariate normal distribution, the symmetry conditions $\Gamma_{ij,p}=\Gamma_{ji,p}$ and $\Gamma_{iL,p}=\Gamma_{Li,p}$ hold for all $i,j\in\mathcal{N}$ (see Proposition~\ref{normal tail exch}). In particular, if $L_i \sim N(\mu_i, \sigma_i^2)$ and $\rho_{ij}$ denotes the Pearson correlation coefficient, then the portfolio return is distributed as
\begin{align*}
	L = \sum_{i=1}^{d}\alpha_iL_i\sim N\left(\sum_{i=1}^{d}\mu_i, \sum_{i=1}^{d}\sum_{j=1}^{d}\alpha_i\alpha_j\sigma_i\sigma_j\rho_{ij}\right).
\end{align*}  Moreover, each pair $(L_i, L)$ forms a bivariate normal random vector.

Given left-tail exchangeability, Equation~\eqref{tail gini} simplifies to
\begin{equation}\label{simp tail gini}
	(\mathrm{TGini}_p(L))^2=\sum_{i=1}^{n}\sum_{j=1}^{n}\alpha_i\alpha_j\Gamma_{ij,p}\mathrm{TGini}_p(L_i)\mathrm{TGini}_p(L_j),
\end{equation} as the asymmetry term $D_{iL, p}=\Gamma_{iL,p}-\Gamma_{Li,p} = 0$ and symmetry $\Gamma_{ij,p}=\Gamma_{ji,p}$ holds. Let $\bm{H}$ denote the $d\times d$ matrix of tail Gini correlation coefficients for the random vector $\bm{L}$, and let $\bm{C}$ be a diagonal $d\times d$ matrix with $\mathrm{TGini}_p(L_i)$ as diagonal entries and zeros elsewhere. We define the MTG analogue of the variance-covariance matrix, $\bm{V}$, as $\bm{V}=\bm{C}\bm{H}\bm{C}.$
Accordingly, the squared tail Gini of the portfolio return is
\begin{equation*}
	(\mathrm{TGini}_p(L))^2 = \bm{\alpha}'\bm{V}\bm{\alpha}.
\end{equation*}
Optimization problem~\eqref{MTG opt prob} is simplified as 
\begin{equation}\label{simp optim problem}
	\begin{aligned}
		\min\  &\bm{\alpha}'\bm{V}\bm{\alpha},\\
		\text{subject to } &\mu = \bm{\alpha}'\bm{\mu},\\
		&1=\bm{\alpha}'\bm{1}.
	\end{aligned}
\end{equation} 
Additional constraints, such as no short selling (i.e., $\alpha_i \geq 0$), can also be incorporated into this formulation. The solution methodology parallels that of the MV optimization framework; see, e.g., \cite{constantinides1995portfolio} and \cite{cai2025conditional}. 

Applying the method of Lagrange multipliers to Problem~\eqref{simp optim problem}, we define the Lagrangian
\begin{equation*}
	L(\alpha, \lambda, \gamma) = \bm{\alpha}'\bm{V}\bm{\alpha} + \lambda (\mu-\bm{\alpha}'\bm{\mu}) +\gamma (1-\bm{\alpha}'\bm{1})
\end{equation*}
where $\lambda$ and $\gamma$ are the Lagrange multipliers. The first-order optimality conditions are:
\begin{equation*}
	\begin{aligned}
		\frac{\partial L(\alpha, \lambda, \gamma)}{\partial \alpha} &= \bm{V}\bm{\alpha} - \lambda \bm{\mu} - \gamma \mathbf{1} = 0, \\
		\frac{\partial L(\alpha, \lambda, \gamma)}{\partial \lambda} &= \mu - \bm{\alpha}' \bm{\mu} = 0, \\
		\frac{\partial L(\alpha, \lambda, \gamma)}{\partial \gamma} &= 1 - \bm{\alpha}' \mathbf{1} = 0,
	\end{aligned}
\end{equation*}
Solving these equations yields:
\begin{equation*}
	\begin{aligned}
		\bm{V}\bm{\alpha} &= \lambda \bm{\mu} + \gamma \mathbf{1} \\
		\mu &= \bm{\alpha}'\bm{\mu}\\
		1&=\bm{\alpha}'\bm{1}, 
	\end{aligned}	
\end{equation*}
Multiplying both sides of $\bm{V}\bm{\alpha} = \lambda \bm{\mu} + \gamma \mathbf{1}$ by $\bm{V}^{-1}$ gives
\begin{equation}\label{alpha_1}
	\bm{\alpha} =  \lambda \bm{V}^{-1} \bm{\mu} + \gamma \bm{V}^{-1} \mathbf{1}
\end{equation}
Substituting this expression into the constraint equations, we obtain:
\begin{equation*}
	\begin{aligned}
		\mu &=\lambda \bm{\mu}' \mathbf{V}^{-1} \bm{\mu} + \gamma \mathbf{1}' \mathbf{V}^{-1} \bm{\mu} \\
		1 &= \lambda \bm{\mu}' \mathbf{V}^{-1} \mathbf{1} + \gamma \mathbf{1}' \mathbf{V}^{-1} \mathbf{1}
	\end{aligned}
\end{equation*}
By solving this system of equations, we have
\begin{equation*}
	\lambda = \frac{C \mu - A}{D}
\end{equation*}
and
\begin{equation*}
	\gamma = \frac{B - A \mu}{D}
\end{equation*}
where $A = \mathbf{1}' \mathbf{V}^{-1} \bm{\mu},$ $B = \bm{\mu}' \mathbf{V}^{-1} \bm{\mu},$ $C = \mathbf{1}' \mathbf{V}^{-1} \mathbf{1},$ and $D = BC - A^2.$ Substituting $\lambda$ and $\gamma$ back into Equation~\eqref{alpha_1}, we obtain the optimal portfolio weights
\begin{equation}\label{optimal weights}
	\bm{\alpha}_p = \mathbf{x} + \mu \mathbf{y},
\end{equation}
where
\begin{equation*}
	\mathbf{x} = \frac{B \mathbf{V}^{-1} \mathbf{1} - A \mathbf{V}^{-1} \bm{\mu}}{D}, \quad \mathbf{y} = \frac{C \mathbf{V}^{-1} \bm{\mu} - A \mathbf{V}^{-1} \mathbf{1}}{D}.
\end{equation*}

Given an expected portfolio return $\mu$ and the mean vector $\bm{\mu}$ calibrated from market data, the efficient frontier can be constructed under the MTG framework using Equation~\eqref{optimal weights}.

It is important to note that this derivation assumes left-tail exchangeability among all return pairs and between each return and the portfolio return. In cases where this assumption does not hold, we employ numerical optimization methods and present empirical analyses in subsequent sections.

\section{Data description and parameter calibration}\label{data desc and params calib}
The parameters of the portfolio’s return distribution, constructed from individual index returns \( L_i \sim F_i \) for \( i = 1, 2, \dots, 6 \), are calibrated using risk measures derived from 1,627 historical daily observations in the U.S.\ securities market, covering the period from April 2018 to September 2024. The portfolio \( L \) comprises six S\&P indices, each representing a distinct asset class: the S\&P 500 (\( L_1 \)), SmallCap 600 (\( L_2 \)), 500 Bond Index (\( L_3 \)), U.S.\ Treasury Bond Index (\( L_4 \)), Cryptocurrency MegaCap Index (\( L_5 \)), and Cryptocurrency LargeCap Ex-MegaCap Index (\( L_6 \)). The corresponding random vector of returns is denoted by \( \bm{L} = (L_1, L_2, \dots, L_6)' \). The return data were obtained from the S\&P Global website. Table~\ref{shorthands} provides a summary of the selected indices, the associated asset classes, and their abbreviations.

The two cryptocurrency indices warrant particular attention. The MegaCap Index comprises only Bitcoin and Ethereum, while the LargeCap Index includes the 60 largest cryptocurrencies by market capitalization, excluding Bitcoin and Ethereum.

\begin{table}[ht]
	\captionsetup{singlelinecheck=false, labelfont=bf}
	\caption{S\&P indices and corresponding assets.}
	\centering
		\begin{tabularx}{\textwidth}{Xll}
			\toprule
			Index                                         & Asset                      & Abbreviation \\  
			\midrule
			S\&P 500                                      & Large-cap stocks           & LCS       \\ 
			S\&P SmallCap 600                             & Small-cap stocks 		   & SCS       \\ 
			S\&P 500 Bond Index                           & Large-cap corporate bonds  & LCB       \\
			S\&P U.S. Treasury Bond Index                 & U.S. treasury bonds        & TB        \\
			S\&P Cryptocurrency MegaCap Index             & Mega-cap cryptocurrencies  & MCC       \\
			S\&P Cryptocurrency LargeCap Index & Large-cap cryptocurrencies & LCC       \\
			\bottomrule
		\end{tabularx}
	\label{shorthands}
\end{table}
In practice, rational investors are especially concerned with downside risk, which quantifies the probability and extent of returns falling below the mean, rather than overall volatility. This concept was introduced by \cite{bawa1977capital} in the development of an alternative capital asset pricing model and has since become a cornerstone of modern portfolio optimization (see \cite{lettau2014conditional} and \cite{zhang2021downside} for further developments). Table~\ref{descriptive} presents the notation and calibrated parameter values for the return distributions used in our empirical analysis. The tail standard deviation of $L_i$ at prudence levels $p$, $\mathrm{SD}_{p}(L_i)$, is defined as the square root of $\mathrm{TV}_{p}(L_i)$. It is evident that the asset rankings based on standard deviation align with those obtained from the GMD and tail Gini measures. This consistency suggests that variation across assets is primarily driven by risk magnitude and cross-asset correlations - a conclusion that aligns with the findings of \cite{shalit2005mean}.

\begin{table}[ht]
	\captionsetup{singlelinecheck=false, labelfont=bf}
	\caption{Descriptive statistics for returns $L_i$, $i=1,2,\dots,6.$ The unit of the calibrated values is percentage.}
	\centering
	\begin{tabularx}{\textwidth}{Xcccccc}
		\toprule
		Asset & Mean $(\mu_i)$ & Std $(\sigma_i)$ & $\mathrm{SD}_{0.10}(L_i)$ & $\mathrm{SD}_{0.05}(L_i)$  & $\mathrm{TGini}_{0.10}(L_i)$ & $\mathrm{TGini}_{0.05}(L_i)$  \\
		\midrule
		LCS   & 0.05523        & 1.25800          & 1.36405                   & 1.58363                    & 1.15597                      & 1.32742                       \\
		SCS   & 0.03817        & 1.61125          & 1.66671                   & 2.00388                    & 1.33174                      & 1.62512                       \\
		LCB   & 0.01302        & 0.38349          & 0.36796                   & 0.44035                    & 0.29917                      & 0.36710                       \\
		TB    & 0.00669        & 0.30274          & 0.19422                   & 0.20851                    & 0.18445                      & 0.19676                       \\
		MCC   & 0.20170        & 4.15189          & 3.44267                   & 3.61466                    & 3.35958                      & 3.65508                       \\
		LCC   & 0.09703        & 4.93953          & 4.13265                   & 4.28499                    & 4.05566                      & 4.34643                       \\
		\bottomrule
	\end{tabularx}\label{descriptive}
\end{table}

To further explore the tail characteristics of crypto-asset returns, we analyze Binance Coin (BNB) and Cardano (ADA), two major constituents of the LargeCap Index. We fit their return series, obtained from Yahoo Finance for the period 2020-2021, to the generalized Pareto distribution. The estimated shape parameters, reported in Table~\ref{GPD parameters}, exceed 0.5 and below 1, indicating infinite tail standard deviations but finite means during this period. We refer readers to \cite{chen2024infinite} for the explanation of infinite second moments like standard deviations of the generalized Pareto distribution. However, the tail Gini measure remains finite, thereby reinforcing its suitability as a robust measure of variability for heavy-tailed assets within the efficient frontier framework.

\begin{table}[ht]
	\centering
	\captionsetup{labelfont=bf}
	\caption{Parameter estimators of the generalized Pareto distribution for BNB and ADA. The values are obtained by fitting the returns data from 2020 through 2021.}
	\begin{tabularx}{200pt}{Xll}
		\toprule
		Asset & BNB & ADA \\
		\midrule
		Prudence level $p$ & 0.10 & 0.05 \\
		\midrule
		Shape & 0.5773 & 0.7079 \\
		Scale & 2.6068 & 2.0407 \\
		\bottomrule
	\end{tabularx}\label{GPD parameters}
\end{table}

Tables~\ref{tail gini corr coeff 0.10} and~\ref{tail gini corr coeff 0.05} report the empirical tail Gini correlation coefficient matrices at prudence levels \( p = 0.10 \) and \( p = 0.05 \), respectively. Importantly, these matrices are not symmetric, indicating that return pairs \( (L_i, L_j) \) for \( i, j = 1, 2, \dots, 6 \), with \( i \neq j \), are not left-tail exchangeable. This asymmetry undermines the common assumption of multivariate Gaussian typically invoked in modern portfolio theory. Consequently, Proposition~\ref{normal tail exch}, which relies on this assumption, is not applicable in practice for optimal portfolio selection. This observation is consistent with a large body of empirical literature documenting the non-normality of individual asset and portfolio return distributions; see, for example, \cite{chunhachinda1997portfolio}, \cite{jondeau2006optimal}, and \cite{adcock2015skewed}.

\begin{table}[ht]
	\captionsetup{labelfont=bf}
	\caption{Tail Gini correlation coefficient matrix of $(L_1,L_2,\dots,L_6)'$ at the prudence level $p=0.10$.}
	\centering
	\begin{tabular}{|c|cccccc|}
		\hline
		\backslashbox{$\Gamma_{ij, 0.10}$}{$\Gamma_{ji, 0.10}$} & LCS & SCS & LCB & TB & MCC & LCC \\
		\hline
		LCS   & 1.0000 & 0.7740 & 0.5824 & 0.4491 & 0.3583 & 0.0668    \\
		SCS   & 0.7951 & 1.0000 & 0.7438 & 0.6963 & 0.4059 & 0.0493    \\
		LCB   & 0.6789 & 0.6506 & 1.0000 & 0.7978 & 0.6708 & 0.8901    \\
		TB    & 0.4203 & 0.5426 & 0.7827 & 1.0000 & 0.5322 & 0.5995    \\
		MCC   & 0.3998 & 0.3666 & 0.6962 & 0.6997 & 1.0000 & 0.6812    \\
		LCC   & 0.3185 & 0.1737 & 0.8695 & 0.4642 & 0.6921 & 1.0000   \\
		\hline
	\end{tabular}\label{tail gini corr coeff 0.10}
\end{table}

\begin{table}[ht]
	\captionsetup{labelfont=bf}
	\caption{Tail Gini correlation coefficient matrix of $(L_1,L_2,\dots,L_6)'$ at the prudence level $p=0.05$.}
	\centering
	\begin{tabular}{|c|cccccc|}
		\hline
		        \backslashbox{$\Gamma_{ij,0.05}$}{$\Gamma_{ji,0.05}$} & LCS & SCS & LCB & TB & MCC & LCC \\
		                 \hline
			LCS    & 1.0000 & 0.7705 & 0.7933 & 0.6359 & 0.2771 & 0.7649    \\
			SCS    & 0.8079 & 1.0000 & 0.8850 & 0.8032 & 0.2031 & 0.7548    \\
			LCB    & 0.8026 & 0.8668 & 1.0000 & 0.8060 & 1.0000 & 0.9881    \\
			TB    & 0.8266 & 0.7853 & 0.7783 & 1.0000 & 0.9304 & 1.0000    \\
			MCC   & 0.4334 & 0.4554 & 1.0000 & 0.8735 & 1.0000 & 0.5853    \\
			LCC   & 0.7782 & 0.6570 & 0.9934 & 1.0000 & 0.5401 & 1.0000    \\
			\hline
	\end{tabular}\label{tail gini corr coeff 0.05}
\end{table}

\section{Empirical MTG efficient frontier}\label{empirical MTG EF}
This section constructs MTG efficient frontiers based on the empirical data described previously, using the algorithm detailed in Appendix~\ref{algorithm for ef}. The efficient frontier is formed by identifying optimal portfolio weights corresponding to a specified set of portfolio's mean targets. For comparative purposes, we also present the classical MV and MTV efficient frontiers to evaluate the relative performance of the MTG approach. 

\subsection{Efficient frontiers with short-selling}
Table~\ref{ef with short sells} presents efficient portfolios constructed from six indices for a range of target returns under the assumption that short selling is permitted. Short selling allows investors to borrow shares of assets that are expected to decline in value.

Panel~A of Table~\ref{ef with short sells} presents the composition of the MV efficient portfolios, with $\sigma_L$ representing the standard deviation of the portfolio. As the target expected return increases, the percentages allocated to LCS, LCB, and MCC also rise, while those for SCS, TB, and LCC decrease. In other words, to achieve a higher expected return, investors should borrow more shares of SCS, TB, and LCC, reallocating capital toward higher-returning assets. This observation is consistent with the fact that SCS and LCC exhibit lower average returns but higher standard deviations compared to LCS and MCC, respectively (see Table~\ref{descriptive}). In addition, the Sharpe ratio for TB is lower than that for LCB, rendering TB less preferable (see \cite{sharpe1966mutual,sharpe1994sharpe} for further details). 

Panel~B of Table~\ref{ef with short sells} illustrates the MTG efficient portfolios at a prudence level of $p = 0.10$. The relationship between expected return and the asset weights - excluding LCB - is consistent with that observed in the MV portfolios; however, the magnitude of the weights differs. For instance, the positive weights in LCS (and the negative shares in SCS) is amplified, while that for TB is mitigated. Consequently, for a given required return, the weights assigned to LCS and TB are higher in the MTG portfolios compared to the MV portfolios, whereas the weights for SCS and LCB tend to be lower. The changes in the weights of MCC and LCC are less pronounced. Notably, the pattern of LCB allocation differs between the MV and MTG frameworks: in the former case, LCB’s weight increases with mean return, whereas in the later it decreases, and the magnitude of this difference is significant. Furthermore, among assets subject to short selling, SCS is utilized more aggressively in the MTG framework, while TB is more prominent in the MV portfolios.

Panel~C of Table~\ref{ef with short sells} presents the MTG efficient portfolios at a prudence level of $p = 0.05$. When compared with the MV efficient portfolios, the conclusions drawn in Panel~B largely persist, though with slight differences in the magnitude. For example, the degree of short selling for SCS is more pronounced in Panel~C than in Panel~B. Panels~D and E of Table~\ref{ef with short sells} present the MTV efficient portfolios at prudence levels of $p = 0.10$ and  $p = 0.05$, respectively, with portfolio weights displaying properties similar to those observed in Panels~B and C.

\begin{landscape}
\begin{table}
	\centering
	\captionsetup{singlelinecheck=false, labelfont=bf}
	\caption{Mean-Variance, Mean-TGini$_p(L)$ and Mean-SD$_p(L)$ efficient frontiers for some selected mean returns with short-selling. The number inside parentheses is the rate of distortion, which is computed using optimal portfolio weights from MTG and MTV with identical mean and prudence level. All values are in percentages.}
	\begin{tabularx}{\linewidth}{@{\extracolsep{\fill}} lccccccccccccc}
		\toprule
		$\mu$ & $\rho(L)$ & LCS & & SCS & & LCB & & TB & & MCC & & LCC & \\
		\midrule
		\multicolumn{14}{l}{Panel A. Mean-Variance ($\rho(L):=\sigma_L$)} \\
		\midrule
		0.10 & 1.23 & 86.78  &         & -50.08  &          & 140.70 &         & -93.01  &          & 41.98  &         & -26.37 &           \\
		0.20 & 2.54 & 170.64 &  & -106.24 & & 343.54 & & -340.99 & & 88.20  & & -55.15 &  \\
		0.30 & 3.85 & 254.50 &  & -162.40 &  & 546.39 &  & -588.98 &  & 134.42 & & -83.93 &   \\
		0.40 & 5.17 & 338.36 &  & -218.56 &  & 749.23 &  & -836.96 & & 180.65 &  & -112.72&  \\
	
		\midrule
		\multicolumn{14}{l}{Panel B. Mean-Tail Gini ($\rho(L):=\mathrm{TGini}_{0.10}(L)$)} \\
		\midrule
		0.10 & 0.84 & 140.24 &         & -85.15  &          & -1.83 &          & 34.17  &           & 38.98  &         & -26.41 &   \\
		0.20 & 1.76 & 292.70 & & -183.65 & & -6.29 & & -28.98 &  & 81.93  & & -55.71 &  \\
		0.30 & 2.69 & 452.30 &  & -307.88 &  & -7.20 &  & -80.89 &  & 125.81 &  & -82.14 &  \\
		0.40 & 3.60 & 587.91 &  & -369.29 &  & -8.92 &  & -162.71&  & 168.97 &  & -115.96&  \\
		\midrule
		\multicolumn{14}{l}{Panel C. Mean-Tail Gini ($\rho(L):=\mathrm{TGini}_{0.05}(L)$)} \\
		\midrule
		0.10 & 0.91 & 169.84 &         & -127.24 &          & 24.13  &          & 15.16  &          & 31.56  &         & -13.44 &         \\
		0.20 & 1.92 & 304.99 &  & -260.98 & & 89.24  &  & -84.36 & & 72.23  & & -21.13 &  \\
		0.30 & 2.80 & 526.13 &  & -362.47 &  & -35.81 & & -76.76 &    & 105.19 &  & -56.28 &  \\
		0.40 & 3.74 & 679.73 &  & -469.09 &  & -44.31 & & -130.22&  & 149.15 &  & -85.26 &  \\
		\midrule
		\multicolumn{14}{l}{Panel D. Mean-Tail Variance ($\rho(L):=\mathrm{SD}_{0.10}(L)$)} \\
		\midrule
		0.10 & 0.89 & 154.98 & (10.51) & -106.10 & (-24.60) & -20.98 & (-1046.45) & 56.50 & (65.35) & 36.99 & (-5.11) & -21.40 & (18.97) \\
		0.20 & 1.85 & 323.04 & (10.37) & -230.83 & (-25.69) & -28.34 & (-350.56) & 2.60 & (108.97) & 77.06 & (-5.94) & -43.52 & (21.88) \\
		0.30 & 2.81 & 483.42 & (6.88) & -340.64 & (-10.64) & -51.17 & (-610.69) & -41.27 & (48.98) & 118.72 & (-5.64) & -69.08 & (15.90) \\
		0.40 & 3.78 & 645.58 & (9.81) & -465.93 & (-26.17) & -27.28 & (-205.83) & -120.86 & (25.72) & 159.04 & (-5.88) & -90.55 & (21.91) \\
		\midrule
		\multicolumn{14}{l}{Panel E. Mean-Tail Variance ($\rho(L):=\mathrm{SD}_{0.05}(L)$)} \\
		\midrule
		0.10 & 0.88 & 182.77 & (7.61) & -135.31 & (-6.34) & -20.10 & (-183.30) & 54.92 & (262.27) & 31.00 & (-1.77) & -13.28 & (1.19) \\
		0.20 & 1.82 & 362.72 & (18.93) & -270.99 & (-3.84) & -17.32 & (-119.41) & -12.89 & (84.72) & 65.81 & (-8.89) & -27.32 & (-29.29) \\
		0.30 & 2.77 & 554.80 & (5.45) & -411.16 & (-13.43) & -37.29 & (-4.13) & -63.59 & (17.16) & 99.44 & (-5.47) & -42.20 & (25.02) \\
		0.40 & 3.72 & 746.68 & (9.85) & -555.71 & (-18.47) & -49.91 & (-12.64) & -117.87 & (9.48) & 133.34 & (-10.60) & -56.52 & (33.71) \\
		\bottomrule
	\end{tabularx}
	\label{ef with short sells}
\end{table}
\end{landscape}

\EF{By comparing results in Panels~B and D (as well as the corresponding Panels~C and E), we confirm our initial motivation that $\mathrm{L}^2$-norm risk measure, such as the tail variance, distorts deviations, and consequently amplifies the optimal portfolio weights. For example, the allocation to LCB decreases nearly tenfold from -1.83\% under the MTG framework (Panel~B) to -20.98\% under MTV (Panel~D), representing a distortion rate of -1046.45\%, at the same prudence level $p=0.10$ and mean return $\mu=0.10\%$. Tail variance tends to magnify positive portfolio weights and further deepen negative ones, as reflected in the distortion rates for LCS, SCS, and LCB in Panel~D and E. This amplification effect is notably less severe for MCC and LCC, likely because their sample tail variances fail to stabilize and may diverge erratically in the presence of infinite-variance return distributions \citep{chen2024infinite}. Table \ref{GPD parameters} has illustrated that cryptocurrency returns exhibit infinite variances but finite means.}

\begin{figure}[ht]
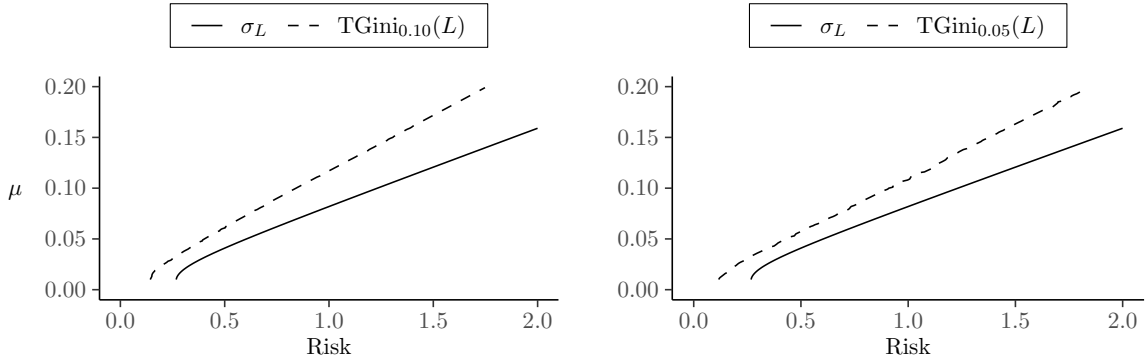

	\begin{subfigure}{0.5\textwidth}
		\input{tg90_sd.tex}
	\end{subfigure}
	\begin{subfigure}{0.5\textwidth}
		\input{tg95_sd.tex}
	\end{subfigure}
	\vspace{-30pt}
	\captionsetup{singlelinecheck=false, labelfont=bf}
	\caption{Efficient frontiers for TGini$_p(L)$ at prudence levels of $p=0.10$ (left) and $p=0.05$ (right), alongside the MV efficient frontier. Short sells are allowed.}
	\label{ef}
\end{figure}

Figure~\ref{ef} depicts the efficient frontiers constructed using the $\mathrm{TGini}_p(L)$ measure at prudence levels of $p=0.10$ (left panel) and $p=0.05$ (right panel), in comparison with those obtained using standard deviation. Notably, the MTG efficient frontier (dashed line) lies above the MV efficient frontier (solid line) in both cases, indicating that for an equivalent level of risk the MTG approach delivers a higher expected return. Furthermore, the difference between the two frameworks becomes more significant as the risk level increases, suggesting that the benefits of employing the $\mathrm{TGini}_p(L)$ measure are amplified under higher risk conditions. However, when transitioning from $p=0.10$ to $p=0.05$, this advantage diminishes slightly (see the first two columns of Panels~B and C in Table 6 for further details).

\subsection{Efficient frontiers without short-selling}
In comparison to Table~\ref{ef with short sells}, Table~\ref{ef with no short-selling} presents the efficient frontiers and portfolio compositions for the MV, MTG, and MTV frameworks - with prudence levels $p=0.10$ and $p=0.05$ - under the assumption that short selling is not permitted. In all efficient portfolios, the proportion allocated to LCS exceeds that of SCS, except in the case of the second lowest expected return in Panels~A and B. Additionally, the weight of MCC increases with the mean across all panels, while the weight of TB decreases.

Furthermore, LCC is consistently excluded from all efficient portfolios. This outcome is expected, given that the mean return of MCC is substantially higher than that of LCC, while its associated risk measures - such as standard deviation and tail Gini - are lower. Overall, the MV efficient portfolios tend to incorporate more assets compared to those constructed using the alternative frameworks for a given required returns. For instance, when $\mu=0.05$, the MV portfolio comprises five assets, whereas the MTG portfolio at $p=0.05$ includes only three. In addition, LCB is incorporated in all portfolios except for the MTV portfolio at $p=0.05$. Notably, when the required average return is low, more than half of the holdings are allocated to LCB and TB; conversely, when the mean exceeds 0.15, the majority of the fund is allocated to MCC.

\begin{table}
	\centering
	\captionsetup{singlelinecheck=false, labelfont=bf}
	\caption{Mean-Variance, Mean-TGini$_p(L)$ and Mean-SD$_p(L)$ efficient frontiers for some selected mean returns without short-selling. All values are in percentages.}
	\begin{tabularx}{\linewidth}{@{\extracolsep{\fill}} lccccccc}
		\toprule
		$\mu$ & $\rho(L)$ & LCS & SCS & LCB & TB & MCC & LCC \\
		\midrule
		\multicolumn{8}{l}{Panel A. Mean-Variance ($\rho(L):=\sigma_L$)} \\
		\midrule
		0.05 & 0.82 & 18.10 & 1.49  & 36.92 & 27.23 & 16.26  &   \\
		0.10 & 1.87 & 14.50 & 14.63 & 15.08 & 14.40 & 41.39  &   \\
		0.15 & 2.97 & 9.02  & 6.48  & 7.57  & 6.97  & 69.95  &   \\
		0.20 & 4.11 & 0.21  & 0.02  & 0.65  & 0.08  & 99.05  &   \\
		\midrule
		\multicolumn{8}{l}{Panel B. Mean-Tail Gini ($\rho(L):=\mathrm{TGini}_{0.10}(L)$)} \\
		\midrule
		0.05 & 0.64 & 21.31 &       & 19.71 &  42.72 & 16.26 &   \\
		0.10 & 1.51 & 14.50 & 14.64 & 15.03 &  14.44 & 41.39 &   \\
		0.15 & 2.41 & 9.02  & 6.48  & 7.57  &  6.97  & 69.95 &   \\
		0.20 & 3.33 & 0.68  & 0.13  & 0.23  &  0.03  & 98.93 &  \\
		\midrule
		\multicolumn{8}{l}{Panel C. Mean-Tail Gini ($\rho(L):=\mathrm{TGini}_{0.05}(L)$)} \\
		\midrule
		0.05 & 0.74 & 19.68 &       &       & 63.01 & 17.31 &    \\
		0.10 & 1.71 & 14.74 & 14.86 & 11.93 & 17.08 & 41.39 &    \\
		0.15 & 2.62 & 9.03  & 7.99  &       & 13.03 & 69.95 &   \\
		0.20 & 3.62 & 1.03  &       &  0.10 &       & 98.87 &    \\
		\midrule
		\multicolumn{8}{l}{Panel D. Mean-Tail Variance ($\rho(L):=\mathrm{SD}_{0.10}(L)$)} \\
		\midrule
		0.05 & 0.72 & 18.48 &      &      & 63.91 & 17.61 &   \\
		0.10 & 1.58 & 23.50 &      & 13.66& 21.29 & 41.55 &    \\
		0.15 & 2.50 & 7.69  & 7.28 & 7.41 & 7.46  & 70.16 &    \\
		0.20 & 3.41 & 0.68  & 0.13 & 0.23 & 0.03  & 98.93 &    \\
		\midrule
		\multicolumn{8}{l}{Panel E. Mean-Tail Variance ($\rho(L):=\mathrm{SD}_{0.05}(L)$)} \\
		\midrule
		0.05 & 0.81 & 14.30 &      &   & 67.05 & 18.65 &    \\
		0.10 & 1.71 & 23.48 &      &   & 34.52 & 42.00 &     \\
		0.15 & 2.63 & 10.34 & 3.85 &   & 15.53 & 70.29 &     \\
		0.20 & 3.58 & 0.63  &      &   & 0.40  & 98.97 &     \\
		\bottomrule
	\end{tabularx}
	\label{ef with no short-selling}
\end{table}

The second panel of Table~\ref{ef with no short-selling} presents the MTG efficient portfolios at a prudence level of $p=0.10$. SCS is included in all portfolios for the specified return levels, except in the lowest return case ($\mu=0.05$). Under MTG with $p=0.10$, the weight assigned to MCC is identical to that in the MV portfolio, except when $\mu=0.2$. LCS is more favorable in the MTG efficient portfolios than in the MV framework, whereas LCB is less preferred. Overall, the number of assets in the MTG portfolios is comparable to that in the MVs when the selected return exceeds 0.05.

The final three panels of Table ~\ref{ef with no short-selling} present the efficient frontiers and portfolio compositions for MTG ($p=0.05$), MTV ($p=0.10$), and MTV ($p=0.05$). The trend toward less number of assets portfolios persists, as SCS and LCB are frequently excluded - particularly in the MTV portfolio at $p=0.05$. Moreover, the weight assigned to MCC remains relatively stable across all frameworks, regardless of the risk measure employed.

\begin{figure}[ht]
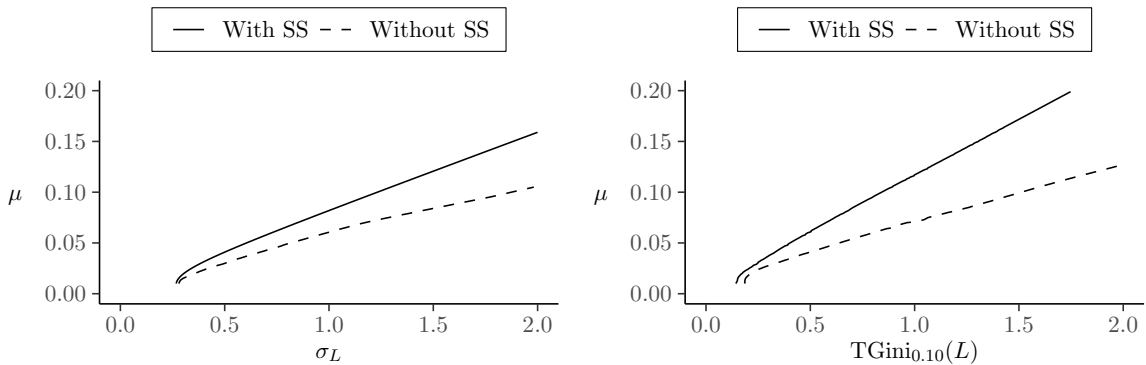

	\begin{subfigure}{0.5\textwidth}
		\input{sd_nss.tex}
	\end{subfigure}
	\begin{subfigure}{0.5\textwidth}
		\input{tg90_nss.tex}
	\end{subfigure}
	\vspace{-30pt}
	\captionsetup{singlelinecheck=false, labelfont=bf}
	\caption{MV (left) and MTG (right) efficient frontiers with and without short-selling. The prudence level for TGini$_p(L)$ in the MTG is equal to $0.10$.}
	\label{effect of short-selling}
\end{figure}

To examine the impact of short selling, Figure~\ref{effect of short-selling} compares the efficient frontiers of MV (left panel) and MTG (right panel, with $p=0.10$) portfolios. In both cases, the efficient frontier obtained with short selling lies above that without short selling, and the disparity increases as the risk measure rises. This observation indicates that short selling enables investors to achieve a higher expected return for a given level of risk, thereby enhancing market efficiency - a finding that is consistent with the literature. For example, \cite{saffi2011price} analyze how short-sale constraints imposed during the 2007-2009 financial crisis affect market efficiency using a global dataset. Additionally, Figure~\ref{effect of short-selling} reveals that short selling can amplify the portfolio’s expected return, so that the mean of an efficient portfolio incorporating short selling may exceed the maximum mean of the six individual assets. For instance, although Table~\ref{descriptive} shows that MCC has a maximum expectation of 0.20170, efficient portfolios with short selling can exhibit a mean exceeding that value.

%


In summary, the numerical experiments presented in this section yield several key conclusions. First, the tail Gini risk measure offers significant benefits in portfolio optimization. Investors who adopt this prudence-based risk measure can achieve superior expectations for a given level of risk compared to the traditional MV approach, thereby underscoring the advantage of using tail Gini over variance as the risk minimization metric.

Second, all efficient frontiers discussed in this section satisfy the necessary conditions for second-degree stochastic dominance (SSD), as evidenced by the decline in the mean minus risk measures with increasing required return. For a detailed discussion on the necessary and sufficient conditions for SSD, see \cite{yitzhaki1982stochastic}.

Third, by incorporating considerations of downside risk, the resulting efficient portfolios tend to be smaller in terms of number of assets. Portfolios based on tail risk measures tend to concentrate allocations in fewer assets, highlighting a trade-off between holding asset numbers and robustness against adverse events.

Fourth, the use of short selling is demonstrated to be an effective tool for enhancing market efficiency. Allowing short sales expands the investment opportunity set, yielding a higher expected return for a given risk level, as demonstrated by efficient frontiers that outperform those without short selling.

Finally, the MTG framework addresses key limitations inherent in strategies based on $\mathrm{L}^2$-norm risk measures, such as MTV. Specifically, the MTV approach tends to amplify optimal portfolio weights, potentially resulting in overly aggressive investment allocations.

\section{Conclusion}\label{conclusion}

In this paper, we introduce the MTG framework as an alternative to the traditional MV approach for constructing efficient frontiers and optimal portfolios. The MTG method offers several advantages over strategies based on $\mathrm{L}^2$-norm risk measures, such as MTV. For example, it requires only a finite first moment and explicitly focuses on downside risk. In the first part of the paper, we derive an analytic MTG efficient frontier under the assumption of left-tail exchangeability among all pairs of constituent asset returns and between each return and the portfolio return where left-tail exchangeability denotes an equal relationship between the tail Gini correlation coefficients of two RVs.

In the second part, we empirically investigate the MTG efficient frontier using daily returns from six indices representing stocks, bonds, and cryptocurrencies obtained from the U.S.\ securities market. Fitting the empirical data to the generalized Pareto distribution reveals tail index estimates consistent with infinite-variance behavior in cryptocurrency returns. Our findings indicate that, when short selling is allowed, the MTG framework outperforms the MV approach by achieving a higher expected return for a given level of risk. Moreover, the MTG approach demonstrates superior performance over the MTV strategy by mitigating the amplification distortions induced by $\mathrm{L}^2$-norm risk measures. Conversely, in the absence of short selling, MV portfolios tend to include more assets than those prudence-based measures. Finally, we demonstrate that short selling is an effective tool for enhancing market efficiency, consistent with existing literature.

\paragraph{Acknowledgments}
The authors gratefully acknowledge funding from the project ``New Order of Risk Management: Theory and Applications in the Era of Systemic Risk''.

\titleformat{\section}[block]{\Large\bfseries}{Appendix \Alph{section}.}{1ex}{}[]
\begin{appendices}
	\section{Proof of Proposition~\ref{basic p of te}}
	\begin{enumerate}[(1)]
			\item This follows directly from Theorem~2.3 of \cite{furman2017beyond}
			\item We refer to \cite{bernard2017value} for the detail of tail comonotonicity. Let $U$ be a uniform distributed RV with domain in $(0, p^\ast]$. If $L_1$ and $L_2$ are tail comonotonic,
			\begin{equation*}
				\begin{aligned}
					\Gamma_{12,p}&=\frac{\mathrm{Cov}[L_1,F_2(L_2)| L_2<l_{p,2}]}{\mathrm{Cov}[L_1,F_1(L_1)| L_1<l_{p,1}]}=\frac{\mathrm{Cov}[F_1^{-1}(U_1),U_2| F_2^{-1}(U_2)<F_2^{-1}(p)]}{\mathrm{Cov}[F_1^{-1}(U_1),U_1| F_1^{-1}(U_1)<F_1^{-1}(p)]}\\
					&=\frac{\mathrm{Cov}[F_1^{-1}(U),U]}{\mathrm{Cov}[F_1^{-1}(U),U]}=1.
				\end{aligned}
			\end{equation*}
			The third equation holds because $L_i$, $i\in\mathcal{N},$ are comonotonic on $\{U_i<p^\ast\}$. Moreover,
			\begin{equation*}
				\Gamma_{21,p}=\frac{\mathrm{Cov}[L_2,F_1(L_1)| L_1<l_{p,1}]}{\mathrm{Cov}[L_2,F_2(L_2)| L_2<l_{p,2}]}=\frac{\mathrm{Cov}[F_2^{-1}(U),U]}{\mathrm{Cov}[F_2^{-1}(U),U]}=1.
			\end{equation*} Hence, $\Gamma_{12,p}=\Gamma_{21,p}=1$ for all $p\in(0, p^\ast]$.
			\item Analogous to (2), we can prove $\Gamma_{12,p}=\Gamma_{21,p}=-1$ for all $p\in(0, p^\ast]$.
		\end{enumerate}
	
	\section{Proof of Proposition~\ref{normal tail exch}}
	Let $Z_1$ and $Z_2$ be two independent standard normally distributed RVs. For any two $L_i$ and $L_j$ in Gaussian random vector and $i\neq j$, we can express
		\begin{equation*}
			\begin{aligned}
				L_i&=\mu_i+\sigma_i Z_1,\\
				L_j&=\mu_j+\sigma_j \left(\rho_{ij}Z_1+\sqrt{1-\rho_{ij}^2}Z_2\right),
			\end{aligned}
		\end{equation*} where $\mu_i$ and $\mu_j$ are their respective means, $\sigma_i$ and $\sigma_j$ are their respective standard deviations and $\rho_{ij}$ is their Pearson correlation coefficient. By assuming $Z=\rho_{ij}Z_1+\sqrt{1-\rho_{ij}^2}Z_2\sim N(0,1)$,
		\begin{equation*}
			\begin{aligned}
				\Gamma_{ij,p}&=\frac{\mathrm{Cov}[L_i,F_j(L_j)| L_j<l_{p,j}]}{\mathrm{Cov}[L_i,F_i(L_i)| L_i<l_{p,i}]}=\frac{\mathrm{Cov}\left[L_i,\Phi(Z)| \mu_j+\sigma_jZ<\mu_j+\sigma_j \Phi^{-1}(p)\right]}{\mathrm{Cov}[L_i,\Phi(Z_1)| \mu_i+\sigma_i Z_1<\mu_i+\sigma_i\Phi^{-1}(p)]}\\
				&=\frac{\sigma_i\mathrm{E}\left[Z_1(\Phi(Z)-\frac{1}{2})| Z< \Phi^{-1}(p)\right]}{\sigma_i\mathrm{E}\left[Z_1(\Phi(Z_1)-\frac{1}{2})| Z_1<\Phi^{-1}(p)\right]}=\frac{\mathrm{E}\left[Z_1(\Phi(Z)-\frac{1}{2})| Z< \Phi^{-1}(p)\right]}{\mathrm{E}\left[Z_1(\Phi(Z_1)-\frac{1}{2})| Z_1<\Phi^{-1}(p)\right]},
			\end{aligned}
		\end{equation*}
		where $\Phi$ is the CDF of a standard normal. Similarly,
		\begin{equation*}
			\Gamma_{ji,p}=\frac{\mathrm{E}\left[Z(\Phi(Z_1)-\frac{1}{2})| Z_1< \Phi^{-1}(p)\right]}{\mathrm{E}\left[Z(\Phi(Z)-\frac{1}{2})| Z<\Phi^{-1}(p)\right]}.
		\end{equation*} The denominators of $\Gamma_{ij,p}$ and $\Gamma_{ji,p}$ are equal because $Z_1$ and $Z$ are both standard normally distributed RVs. Moreover,
		\begin{equation*}
			\begin{aligned}
				\mathrm{E}\left[\left.Z_1\left(\Phi(Z)-\frac{1}{2}\right)\right\vert Z< \Phi^{-1}(p)\right]&=\int_{-\infty}^{\Phi^{-1}(p)}	\mathrm{E}\left[\left.Z_1\left(\Phi(Z)-\frac{1}{2}\right)\right\vert Z=z\right]f_Z(z)dz\\
				&=\int_{-\infty}^{\Phi^{-1}(p)}\int_{-\infty}^{+\infty}	z_1\left(\Phi(z)-\frac{1}{2}\right)f_{Z_1\vert Z}(z_1\vert z)dz_1f_Z(z)dz \\
				&=\int_{-\infty}^{\Phi^{-1}(p)}\int_{-\infty}^{+\infty}	z_1\left(\Phi(z)-\frac{1}{2}\right)f_{Z_1, Z}(z_1, z)dz_1dz,
			\end{aligned}
		\end{equation*} and 
		\begin{equation*}
			\begin{aligned}
				\mathrm{E}\left[\left.Z\left(\Phi(Z_1)-\frac{1}{2}\right)\right\vert Z_1< \Phi^{-1}(p)\right]&=\int_{-\infty}^{\Phi^{-1}(p)}\int_{-\infty}^{+\infty}	z\left(\Phi(z_1)-\frac{1}{2}\right)f_{Z_1, Z}(z_1, z)dzdz_1,
			\end{aligned}
		\end{equation*} where $f_Z$ is the probability density function of $Z$ and $f_{Z_1, Z}$ is the joint probability density function of a bivariate normally distributed random vector with zero means and unit variances. These two expectations show that the numerators of $\Gamma_{ij,p}$ and $\Gamma_{ji,p}$ are also equal. In summary, $\Gamma_{ij,p}$ and $\Gamma_{ji,p}$ are equal when $i\neq j$. When $i=j$, it is evident that $\Gamma_{ij,p}=\Gamma_{ji,p}=1$.
		
	\section{Proof of Proposition~\ref{tail gini equalvalent to gini}}
	For simplicity, we first consider a portfolio consisting of two returns. Using the properties of covariance and Equations~\eqref{tail gini eq}, we derive that
		\begin{equation*}
			\begin{aligned}
				\mathrm{TGini}_p(L)&=\frac{4}{p} \left\{\alpha_1\mathrm{Cov}[L_1, F(L) | L<l_p]+\alpha_2\mathrm{Cov}[L_2, F(L) | L<l_p]\right\} \\
				&=\alpha_1\Gamma_{1L,p}\mathrm{TGini}_p(L_1)+\alpha_2\Gamma_{2L,p}\mathrm{TGini}_p(L_2).
			\end{aligned}
		\end{equation*} Applying the identity $\Gamma_{iL,p}=\Gamma_{Li,p}+D_{iL, p}$, we obtain
		\begin{equation*}
			\begin{aligned}
				\mathrm{TGini}_p(L)=&\alpha_1(\Gamma_{L1,p}+D_{1L, p})\mathrm{TGini}_p(L_1)+\alpha_2(\Gamma_{L2,p}+D_{2L, p})\mathrm{TGini}_p(L_2)\\
				=&\sum_{i=1}^{2}\alpha_i\Gamma_{Li,p}\mathrm{TGini}_p(L_i)+\sum_{i=1}^{2}\alpha_iD_{iL, p}\mathrm{TGini}_p(L_i)\\
				=&\frac{\sum_{i=1}^{2}\alpha_i\mathrm{TGini}_p(L_i)(\alpha_1\mathrm{TGini}_p(L_1)\Gamma_{1i,p}+\alpha_2\mathrm{TGini}_p(L_2)\Gamma_{2i,p})}{\mathrm{TGini}_p(L)}\\
				&+\sum_{i=1}^{2}\alpha_iD_{iL, p}\mathrm{TGini}_p(L_i)\\
				=&\frac{\alpha_1\alpha_2\mathrm{TGini}_p(L_1)\mathrm{TGini}_p(L_2)(\Gamma_{12,p}+\Gamma_{21,p})+\sum_{i=1}^{2}\alpha_i^2\mathrm{TGini}_p^2(L_i)}{\mathrm{TGini}_p(L)}\\
				&+\sum_{i=1}^{2}\alpha_iD_{iL, p}\mathrm{TGini}_p(L_i).
			\end{aligned}
		\end{equation*} The third equality in the preceding expression holds because 
		\begin{equation*}
			\begin{aligned}
				\Gamma_{Li,p}&=\frac{\mathrm{Cov}[L,F_i(L_i)| L_i<l_{p,i}]}{\mathrm{Cov}[L,F(L)| L<l_p]}=\frac{\frac{4}{p}\sum_{j=1}^{2}\alpha_j\mathrm{Cov}[L_j,F_i(L_i)| L_i<l_{p,i}]}{\mathrm{TGini}_p(L)}\\
				&=\frac{\alpha_1\mathrm{TGini}_p(L_1)\Gamma_{1i,p}+\alpha_2\mathrm{TGini}_p(L_2)\Gamma_{2i,p}}{\mathrm{TGini}_p(L)},
			\end{aligned}
		\end{equation*} while the last equality follows from $\Gamma_{11,p}=\Gamma_{22,p}=1$. Multiplying both sides by $\mathrm{TGini}_p(L)$ and rearranging terms yields
		\begin{equation*}
			\begin{aligned}
				(\mathrm{TGini}_p(L))^2&-\mathrm{TGini}_p(L)\sum_{i=1}^{2}\alpha_iD_{iL, p}\mathrm{TGini}_p(L_i)\\
				&=\alpha_1\alpha_2\mathrm{TGini}_p(L_1)\mathrm{TGini}_p(L_2)(\Gamma_{12,p}+\Gamma_{21,p})+\sum_{i=1}^{2}\alpha_i^2\mathrm{TGini}_p^2(L_i)\\
				&=\frac{1}{2}\sum_{i=1}^{2}\sum_{j=1}^{2}\alpha_i\alpha_j\mathrm{TGini}_p(L_i)\mathrm{TGini}_p(L_j)(\Gamma_{ij,p}+\Gamma_{ji,p}).
			\end{aligned}
		\end{equation*} When $p$ goes to one, the tail Gini of a RV $Y\sim F$ converges to
		\begin{equation*}
			\lim\limits_{p\rightarrow 1}\mathrm{TGini}_p(Y)=4 \mathrm{Cov}[Y, F_Y(Y)]=G_Y.
		\end{equation*} Hence, Equation~\eqref{gini} is obtained by substituting the tail Gini and tail Gini correlations with GMD and Gini correlations respectively in Equation~\eqref{tail gini}.
		
		The generalization to a $d$-dimensional portfolio follows the same principle.
	
	\section{Algorithm for MTG efficient frontiers}\label{algorithm for ef}
	For numerical implementation, \citet{yitzhaki2003gini} propose a computationally efficient formulation of the sample GMD, which improves upon the original expression involving absolute differences. In our analysis, we compute $\mathrm{TGini}_p(L)$ by applying Proposition~\ref{num method} to the subset of observations below the empirical $\mathrm{VaR}_p(L)$.
	
	\begin{proposition}\label{num method}
		
		The GMD in the sample given ordered observations $x_{(1)}\leq x_{(2)}\leq\ldots\leq x_{(n)}$ is given by
		
		\begin{equation}
			G_L = \frac{2}{n(n-1)}\sum_{i=1}^{n-1}x_{(i)}[2i-(n+1)].
		\end{equation}
		
	\end{proposition}
	
	\begin{proof}
		
		We have that
		\begin{align*}
			G_L &= \frac{1}{n(n-1)}\sum_{i=1}^{n}\sum_{j=1}^{n}|x_{(i)}-x_{(j)}|\\
			&= \frac{2}{n(n-1)}\sum_{i=1}^{n-1}\sum_{j>i}^{n}\left(x_{(j)}-x_{(i)}\right)\\
			&= \frac{2}{n(n-1)}\left(\sum_{i=1}^{n-1}-(n-i)x_{(i)}+\sum_{i=1}^{n-1}\sum_{j>i}^{n}x_{(j)}\right)\\
			&= \frac{2}{n(n-1)}\left(\sum_{i=1}^{n-1}-(n-i)x_{(i)}+\sum_{i=1}^{n-1}\sum_{k=1}^{n-i}x_{(i+k)}\right)\\
			&= \frac{2}{n(n-1)}\left[\left(\sum_{i=1}^{n-1}-(n-i)x_{(i)}\right)+x_{(2)}+2x_{(3)}+3x_{(4)}\cdots+(n-1)x_{(n)}\right]\\
			&= \frac{2}{n(n-1)}\left(\sum_{i=1}^{n-1}-(n-i)x_{(i)}+\sum_{i=1}^{n-1}(i-1)x_{(i)}\right)\\
			&= \frac{2}{n(n-1)}\sum_{i=1}^{n-1}x_{(i)}[2i-(n+1)].
		\end{align*}
	\end{proof}  

	Subsequently, we present an algorithm for identifying the MTG efficient portfolio that achieves a specified target mean return.
	
	\begin{algorithm}[H]
		\caption{Find efficient weights $w=(w_1,w_2,\ldots,w_d)'$ given $\bm{L}=(L_1,L_2,\ldots,L_d)$ where $L_i=(x_{i1},x_{i2},\ldots,x_{in})'$ is a column vector of historical returns for asset $i$ and $mt$ is the target mean return of the portfolio. The efficient weights minimize variability measure $f(w,L)$.}
		
		$\mu \gets \textbf{colMeans}(-\bm{L})$;
		$\delta \gets 0.001$;
		$dw \gets \textbf{rep}(0,d)$;
		$dwi \gets \textbf{rep}(0,d)$;
		
		\For{$i$ in $1:d$} {
			\uIf{$i=d$}{
				$i1 \gets 1$;
			}\Else{$i1 \gets i+1$;}
			$ii \gets c(i,i1)$; $dwi[i] \gets \delta$\;
			$dwi[i1] \gets \delta*(sum(\mu[-ii])/(d-2)-\mu[i])/(\mu[i1]-sum(\mu[-ii])/(d-2))$\;
			$dwi[-ii] \gets -(\delta+dwi[i1])/(d-2)$;
			$dw \gets \textbf{rbind}(dw, dwi, -dwi)$;
		}
		$n\_dw \gets nrow(dw)$\;
		
		$w \gets \textbf{rep}(0,d)$;
		$w[1] \gets 1$; $w[2] \gets (mt-\mu[1])/(\mu[2]-sum(\mu[-1,-2])/(d-2))$\;
		$w[-1,-2] \gets $

		$thresh \gets 1$;
		$wf\_o \gets 0$;
		$wf\_n \gets 0$;
		
		\While{$thresh>0.00001$}{
			$w1 \gets \textbf{do.call}(\textbf{rbind},
			\textbf{replicate}(ndw, w, simplify=F)) + dw$;
			$wf\_o \gets wf\_n$;
			$wf \gets c()$;
			
			\For{$i$ in $1:ndw$}{
				$wf \gets c(wf, f(w,L))$;
			}
			
			$w\_min \gets \textbf{which.min}(wf)$
			$w \gets w1[w\_min,]$;
			$wf\_n \gets wf[w\_min]$;
			$thresh \gets |wf\_n-wf\_o|$;
		}
	\end{algorithm}
\end{appendices}

\bibliographystyle{chicago}
\bibliography{Bibliography}
\end{document}